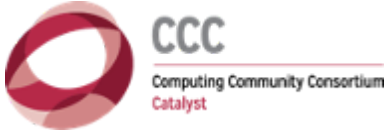

# The Rise of AI-Driven Simulators: Building a New Crystal Ball
*A Computing Community Consortium (CCC) Quadrennial Paper*

*Ian Foster (University of Chicago), David Parkes (Harvard University), and Stephan Zheng (Salesforce AI Research)*

Fifty years ago, weather forecasters struggled to do better than predicting that tomorrow's weather would be the same as today. Today, weather forecasts are often accurate a week or more into the future, allowing individuals and societies to prepare for what is no longer unforeseen. This remarkable transformation is due in large parts to computers, and in particular to the rise of computational simulation, a method for using computers to predict the future state of complex systems.

First developed in the last days of World War II for military purposes, simulation is now pervasive throughout human society and the economy, providing decision makers with a remarkable crystal ball not just for next week's weather but also for how an airplane will perform when it flies through different weather patterns, the effectiveness of a new drug against a new disease, and the behavior of a new material in future batteries.

*Computer simulation is the process of mathematical modelling, performed on a computer, which is designed to predict the behaviour of or the outcome of a real-world or physical system.*[1] A simulation is often configured by dividing space (e.g., North America) into a number of small cells that each hold a set of values (e.g., temperature and pressure) together with a set of local rules for how to update the cells for the next time step (e.g., temperature/pressure a minute from now based on current temperatures and pressures of the cell and neighboring cells). A simulation run is seeded with measured inputs (temperature/pressure) and repeatedly applies its rules to update the simulated system over time. More accurate input data, smaller cells, and better rules enable higher fidelity simulation (e.g., good weather forecasts for next week instead of just tomorrow).

The use of computational simulation is by now so pervasive in society that it is no exaggeration to say that continued U.S. and international prosperity, security, and health depend in part on continued improvements in simulation capabilities. *What if we could predict weather two weeks out, guide the design of new drugs for new viral diseases, or manage new manufacturing processes that cut production costs and times by an order of magnitude? What if we could predict collective human behavior, for example, response to an evacuation request during a natural disaster, or labor response to fiscal stimulus?* (See also the companion CCC Quad Paper on Pandemic Informatics, which discusses features that would be essential to solving large-scale problems like preparation for, and response to, the inevitable next pandemic.)

---

[1] https://en.wikipedia.org/wiki/Computer_simulation

The past decade has brought remarkable advances in complementary areas: in *sensors*, which can now capture enormous amounts of data about the world, and in *AI methods* capable of learning to extract predictive patterns from those data. These advances may lead to a new era in computational simulation, in which sensors of many kinds are used to produce vast quantities of data, AI methods identify patterns in those data, and new AI-driven simulators combine machine-learned and mathematical rules to make accurate and actionable predictions. At the same time, there are new challenges---computers in some important regards are no longer getting faster, and in some areas we are reaching the limits of mathematical understanding, or at least of our ability to translate mathematical understanding into efficient simulation.

In this paper, we lay out some themes that we envision forming part of a cohesive, multi-disciplinary, and application-inspired research agenda on AI-driven simulators. The new opportunity presented through AI driven simulators is to learn from data, to accelerate simulation through prediction, and to augment physics-based simulation with predictive models of social and economic phenomena.

**AI-driven modeling of nonlinear dynamical systems.** For systems that follow the laws of physics, AI techniques are increasingly used to replace steps in expensive simulations, with embedded machine learning models providing fast surrogates. Deep learning is also increasingly used to provide end-to-end models, and replace numerical computation.

AI methods are also used to interpret the results of simulations, and control which computations to perform next. For example, next-generation climate models may use deep learning to fit parametrizations of equation-based, computational models. In earthquake modeling, deep learning can be used to predict the spatial patterns of aftershocks, and associate multi-sensor measurements for the purpose of earthquake detection.

Many challenges remain, including incorporating physical laws into machine learning models in order to make them more accurate and robust (see, too, the companion CCC Quad Paper on Next Wave Artificial Intelligence); combining black-box models of nonlinear dynamical systems with theoretically sound controllers; developing distributed computing architectures for scaling up AI-driven simulations; developing neural network methods for non-Euclidean data such as graph data that reflects traffic flow; and the need for robust machine learning that transfers to new settings and supports uncertainty quantification.[2]

---

[2] See the discussion in the report from a 2018 NITRD workshop on "The Convergence of High Performance Computing, Big Data, and Machine Learning" https://www.nitrd.gov/nitrdgroups/index.php?title=HPC-BD-Convergence For additional, general discussions on the role of AI in data-driven simulations, see Bergen, Karianne J., Paul A. Johnson, Maarten V. de Hoop, and Gregory C. Beroza, Machine learning for data-driven discovery in solid Earth geoscience. *Science* 363 (6433), 2019; Dulac-Arnold, Gabriel, Daniel Mankowitz, and Todd Hester, Challenges of Real-World Reinforcement Learning, *CoRR abs/1904.12901*, 2019; and Schneider, Tapio, Shiwei Lan, Andrew Stuart, and João Teixeira, Earth System

**Modeling multi-agent behavior in complex systems.** There are many open questions and challenges in regard to the use of AI to simulate and model systems that relate to human behavior, for example economic and social systems, bring a new set of challenges. Not least, the ability to simulate and model brings to the forefront ethical questions in regard to control, influence, and the potential for discrrimination.

Within economics, it has been common to make use of equilibrium models to represent the decisions of firms and consumers. These methods are complex and do not readily scale up. Agent-based models are also used, with rule-based simulators to describe different kinds of behaviors and to facilitate "bottom-up" simulations of economic systems. A challenge is that these rules that guide behavior are *ad hoc* and, at the same time, make the results of the simulation hard to interpret.

The use of AI and machine learning may present a new way to model economic and social systems, with AI-based models of different kinds of entities (e.g., consumers, workers, firms, governments) and different kinds of behaviors (e.g., cooperation, coordination, alliances, self-interest). Analogous to the dramatic advances in modeling human language, can there be dramatic advances in modeling human behavior? Could we get closer to simulating an entire economic system? What can be learned about human behavior from large data sets?

**Optimizing decision-making with large-scale reinforcement learning.** Reinforcement learning (RL) is a decision-making learning framework to learn optimal behavior in complex, uncertain environments--- a decision policy that maximizes expected, discounted reward. DeepMind has demonstrated remarkable success in achieving superhuman performance through self-play in games of increasing complexity.[3] These systems learn through access to a game simulator and by self-play, sometimes requiring multiple days of computation to attain superhuman performance.

There remain many challenges in extending RL methods to real-world open systems; e.g., handling high dimensional, continuous action spaces; reasoning about safety conditions; and handling non-stationarity and poorly specified reward functions. One problem is to build accurate enough real-world simulators, such as those used today in the push towards automated vehicles. This can make it necessary to train RL models offline, from the data recovered from an existing, real-world policy. There is also the "sim2real problem": how do you know whether your policies, learned in simulation, will work in the real world (e.g., in application to synthetic drug design, or controllers for commercial jet planes)? Are models accurate enough that RL can be used to make decisions about which costly, real-world experiments to run next, for example which drugs to synthesize and test, which new materials to synthesize and study?

---

Modeling 2.0: A Blueprint for Models That Learn From Observations and Targeted High-Resolution Simulations, *Geophysical Research Letters* 44:12,396-12,417, 2017.

[3] DeepMind's AlphaZero provides super-human ability in multiple different games, including Go and chess, and works without making use of expert data or any in-built domain knowledge except for the basic rules of the game. DeepMind's AlphaStar plays StarCraft II using a deep neural network that is trained directly from raw game data.

Large-scale, multi-agent RL may also suggest a path towards new approaches to economic policy design, such as recent work on the use of multi-agent RL to study taxation policy. But any optimization activity must be done together with a careful examination of the appropriate end states, or valued goals around which the system is oriented. What are the valued ends, who do they serve, do they serve all interests equitably, do they serve societal values, generally? There also remain severe technical challenges around, for example, scaling up from tens of agents to thousands or even millions, introducing behavioral considerations, and calibrating models against real-world data.

A *cross-cutting issue* is that AI-driven simulators need access to high quality data, whether about the economy, weather, traffic patterns, consumption patterns, the flow of information, or earthquakes. The data needs to be representative, available in sufficient quantity, and granular enough. Progress will require the right kind of investment in data, this occurring even while we do not yet have the models to make it clear how this kind of data would be used and valuable.

In summary, AI simulations driven by large-scale, properly curated datasets have significant potential to unlock and accelerate innovation in key areas in science and technology, and across society as a whole. Major recommended areas of investment and research include: acquisition and proper curation of domain-specific data, efficient simulation implementations, focused large-scale hardware investments, fundamental research on improving AI models and algorithms, and attention to the ethical and critical thinking that must go hand-in-hand with technical advances around AI driven simulation and prediction. (See also the companion CCC Quad Paper on Interdisciplinary Approaches to Understanding Artificial Intelligence's Impact on Society.)

*This white paper is part of a series of papers compiled every four years by the CCC Council and members of the computing research community to inform policymakers, community members and the public on important research opportunities in areas of national priority. The topics chosen represent areas of pressing national need spanning various subdisciplines of the computing research field. The white papers attempt to portray a comprehensive picture of the computing research field detailing potential research directions, challenges and recommendations.*

*This material is based upon work supported by the National Science Foundation under Grant No. 1734706. Any opinions, findings, and conclusions or recommendations expressed in this material are those of the authors and do not necessarily reflect the views of the National Science Foundation.*

*For citation use: Foster I., Parkes D., & Zheng S. (2020) The Rise of AI-Driven Simulators: Building a New Crystal Ball.*
*https://cra.org/ccc/resources/ccc-led-whitepapers/#2020-quadrennial-papers*